\documentclass[showpacs, pra,twocolumn,preprintnumbers ,amsmath, amssymb, superscriptaddress, aps]{revtex4-2}
\usepackage{color}
\usepackage{amsmath,amssymb}
\usepackage{pifont}
\usepackage{amssymb}  
\usepackage{bbold}
\usepackage{float}
\usepackage{subfloat} 

\usepackage[caption=false]{subfig}
\usepackage{tikz}
\usepackage{makecell}
\usepackage{subfig}
\usepackage{pifont}   
\usepackage{graphicx} 
\graphicspath{{Figures/}}
\usepackage{dcolumn}  
\usepackage{bm}       
\usepackage{multirow} 
\usepackage{placeins}
\usepackage[colorlinks]{hyperref}
\usepackage{mathtools}

\newcommand{\vect}[1]{\mathbf{#1}}

\begin{document}
	
\title{Semiclassical perspective on Landau levels and Hall conductivity in an anisotropic Cubic Dirac Semi-Metal and the peculiar case of star-shaped classical orbits}
	\date{\today}

	\author{Ahmed Jellal}
	\affiliation{Laboratory of Theoretical Physics, Faculty of Sciences, Choua\"ib Doukkali University, PO Box 20, 24000 El Jadida, Morocco}
	\affiliation{Canadian Quantum  Research Center,
		204-3002 32 Ave Vernon,  BC V1T 2L7,  Canada}
		\author{Hocine Bahlouli}
	\affiliation{Physics Department,  King Fahd University
of Petroleum $\&$ Minerals,
Dhahran 31261, Saudi Arabia}
\author{Michael Vogl}
	\affiliation{Physics Department,  King Fahd University
of Petroleum $\&$ Minerals,
Dhahran 31261, Saudi Arabia}
 \affiliation{Interdisciplinary Research Center (IRC) for Intelligent Secure Systems$,$ KFUPM$,$ Dhahran$,$ Saudi Arabia}

			\begin{abstract}
	We study an anisotropic cubic Dirac semi-metal subjected to a constant magnetic field. In the case of an isotropic dispersion in the $x$-$y$ plane, with parameters $v_{x}=v_{y}$, it is possible to find exact Landau levels, indexed by the quantum number $n$, using the typical ladder operator approach. Interestingly, we find that the lowest energy level (the zero energy state in the case $k_z=0$) has a degeneracy that is three times that of other states. This degeneracy manifests in the Hall conductivity as a step at zero chemical potential that is 3/2 the size of other steps. Moreover, as $n\to\infty$ we find energies $E_n\propto n^{3/2}$, which means the $n$-th step as a function of chemical potential roughly occurs at a value $\mu\propto n^{3/2}$. We propose that these exciting features could be used to identify cubic Dirac semi-metals experimentally.
	Subsequently, we analyze the anisotropic case $v_{y}=\lambda v_{x}$ with $\lambda\neq 1$. First, we consider a perturbative treatment around $\lambda\approx 1$ and find that energies $E_n\propto n^{3/2}$ still holds as $n\to\infty$. To gain further insight into the Landau level structure for a maximum anisotropy, we turn to a semi-classical treatment that reveals interesting star-shaped orbits in phase space that close at infinity. This property is a manifestation of weakly localized states. Despite being infinite in length, these orbits enclose a finite phase space volume and permit finding a simple semi-classical formula for the energy, which again has the form as above. 
 Our findings suggest that both isotropic and anisotropic cubic Dirac semi-metals should leave similar experimental imprints.

			\end{abstract}
	
	\maketitle
	

	\section{ Introduction}
Semi-classical approximations typically are achieved by taking the $\hbar\rightarrow0$ limit of quantum mechanics \cite{10.1063/1.1665596,Kramers1926}. More precisely, this is the limit at which the characteristic action $S$ is large compared to the reduced Planck constant $\hbar$, that is $S\gg\hbar$ \cite{vogl2017semiclassics,loewe2017bridges}. Within this limit, capturing much of the rich and exotic behavior of quantum systems and preserving the intuitive explanations that classical mechanics provides is possible. More than that, semi-classical physics has allowed us to gain deep insights into the interplay between classical chaos and quantum mechanics \cite{ChaosBook}. A prototypical example of this is the effect of quantum scarring that occurs with wavefunctions in quantum billiards that can be traced back to the stable orbits in a corresponding classical chaotic system, where such orbits leave their imprints on the quantum wavefunction \cite{PhysRevLett.53.1515}. An even more important reason for the interest in semi-classical methods is that they provide a non-perturbative approximation scheme, which is analytically accessible - something that is exceptional. 

This strength of the approach can be seen most lucidly in that the semi-classical method can solve problems approximately that ordinary methods such as perturbation theory cannot. For instance, if one treats the harmonic oscillator potential as a perturbation around the free particle, then perturbation theory cannot produce the well-known discrete energy levels. However, in this case, the semi-classical method is impressive because it produces exact results for the discrete energy levels \cite{kleinertsemiclassical,amiot2006mecanique}. 

This work aims to apply semi-classical methods not to the ordinary Schr\"odinger equation with a quadratic dispersion, whose implications are well-known. Instead, we want to see its consequences in the context of more exotic phases of matter - so-called semi-metals.

After the discovery of graphene \cite{doi:10.1126/science.1102896}  (a single layer of carbon atoms arranged in a honeycomb pattern), researchers discovered many exciting transport properties. A famous example is Klein tunneling \cite{Katsnelson2006} showing that an electron under normal incidence will tunnel through a barrier with 100\% probability. These features can be traced back to graphene's peculiar band structure, where the conduction and valence bands touch at isolated points. This property made graphene an early example of a semi-metal -i.e., a material where the overlap between conduction and valence bands is not precisely zero like in an insulator or semiconductor and not of finite measure like in a conductor. Instead, while the overlap is non-zero, it has zero measure. Motivated by this discovery, much interest has emerged in similar classes of materials. One example is Dirac semi-metals \cite{liu2014discovery,liu2017predicted}, which were the first experimentally confirmed material in this class. Dirac semi-metals can be considered a three-dimensional analog of graphene \cite{liu2014discovery,liu2014stable} because, much like graphene, it has a linear dispersion near isolated band touching points. Dirac semi-metals are interesting because the exciting linear band touching is robust. Indeed, Dirac cones do not appear due to accidental band touchings; instead, they are protected by crystalline symmetry and spin-orbit coupling \cite{kushwaha2015bulk,han2020nonsymmorphic} and are therefore very stable. Of course, there is a plethora of other kinds of semi-metals with robust band features, such as nodal line semi-metals where bands touch along a one-dimensional line \cite{fang2016topological,Yu2017} or also semi-metals with band touchings of higher than linear order - such as quadratic or cubic dispersion \cite{Fang_2012,liu2017predicted,Bouhlal_2021,Bouhlal_2022,PhysRevB.105.035141}.
 
In this work, we will focus our investigation on the exotic case of an anisotropic cubic Dirac metal \cite{liu2017predicted,Bouhlal_2021} subjected to a constant magnetic field. We will do so by using both fully quantum and semi-classical methods. Semi-classical results are compared to fully quantum results. Both approaches offer exciting insights and permit us to speculate about how one might experimentally identify a cubic Dirac semi-metal, whether it is isotropic or anisotropic. 

We organized our work as follows. In section \ref{pp1}, we introduce the Hamiltonian of the system. In section \ref{pp2}, we study the Hamiltonian for two cases that allow for an exact quantum mechanical analytical solution - these cases are isotropic in the $x$-$y$ plane. First, we consider the purely two-dimensional limit. Here, the material is thin, and we may restrict ourselves to momentum $k_{z} = 0$ in the $z$-direction for low energies. Next, we consider the case $k_{z}\neq0$, where the material is thick enough so that not all momenta in the $z-$ direction are frozen out. In section \ref{pp3}, we consider perturbatively the effect of the anisotropy by introducing a small anisotropy parameter $\lambda\ll 1$. In section \ref{pp4}, we recast the problem in a semi-classical language, which allows us to investigate classical orbits and study the differences between exact quantum and semi-classical results. 
 We study the maximally anisotropic limit, which is inaccessible to a fully quantum analytical treatment. We find classical orbits and solutions for energy levels that are exact in the semi-classical limit and compare them to numerical results. Lastly, in Sec. \ref{sec:conclusion} we draw our conclusions

	\section{Anisotropic cubic Dirac semi-metals}\label{pp1}

We take the  Hamiltonian describing an anisotropic cubic Dirac semi-metal \cite{liu2017predicted}
	\begin{equation}
		H(k)=\hbar\left( \begin{array}{cc} h(k)&0\\ 0&-h(k) \end{array}\right)
		\label{eq:hfull},
	\end{equation}
	as the starting point of our discussion. Such a Hamiltonian, for instance, is expected to be realizable in quasi-one-dimensional molybdenum
monochalcogenide compounds \cite{liu2017predicted}. For such a material the basis set that was used to write the above Hamiltonian is  $|{\Psi}\rangle=(| A,\uparrow\rangle,| B,\uparrow\rangle,| A,\downarrow\rangle,| B,\downarrow\rangle)$, with $A/B$ and $\uparrow/\downarrow$  designating the orbital and spin degrees of freedom, respectively. Of course, there is the possibility that similar Hamiltonians may be realizable in other types of materials and also with different degrees of freedom (such as sublattice degrees of freedom), giving rise to an identical matrix structure. For this reason, our results remain general in what follows, so we will not explicitly use the basis set introduced above.
	
The Hamiltonian \eqref{eq:hfull} for our case of a cubic Dirac semimetal consists of two blocks with the form
	\begin{equation}
		h(k)=v_{x}(\hat{k}^{3}_{+}+\hat{k}^{3}_{-})\sigma_{x}	+iv_{y}(\hat{k}^{3}_{+}-\hat{k}^{3}_{-})\sigma_{y}+v_{z}k_{z}\sigma_{z} 
		\label{eq:hupperblock},
	\end{equation}
where
	$\hat{k}_{\pm}=\hat{k}_{x}\pm i\hat{k}_{y}$ are momentum operators with $\hat{k}_{i}=	-i\hbar \partial_{i}$. Finally, $\sigma_{i}$ are Pauli matrices, and $v_{x,y,z}$ are real-valued and independent coefficients (they are not velocities) with dimension $T^{-1}$, $T$ stands for time. 
	
	A constant magnetic field $B$ can be introduced into our description with a convenient choice of the Landau gauge. That is, it is introduced via a vector potential $\vect{A}=(0,Bx,0)$ using the minimal substitution procedure $ {\vect k}\to  {\vect k}-\frac{e}{\hbar c}\vect{A}$.  In the unit system ($e=c=\hbar=1$), we therefore replace canonical momentum $\hat p_i$ operators by kinetic momentum operators $\hat \pi_i$ as indicated below	
	\begin{equation}
		\hat{k}_{\pm}\to\hat \pi_{\pm}=\hat{k}_{x}\pm i(\hat{k}_{y}-B\hat x).
	\end{equation}
	
	In the following section, we will first study this Hamiltonian's energy levels - so-called Landau levels - fully quantum mechanically, which is only possible in specific high symmetry situations.
	
	\section{Exact energy levels for the isotropic case \texorpdfstring{$v_{x}=v_{y}$}{va=v} }\label{pp2}
	\subsection{Isotropic case with  \texorpdfstring{$k_{z} = 0 $}{kz=0} }
	The Hamiltonian describing an isotropic cubic Dirac semi-metal with $v_{x}=v_{y}=v$ and the $z$-direction frozen out as $k_{z} = 0 $ (see the appendix of \cite{Bouhlal_2021} for a more detailed discussion on how this happens for a thin material), is given by 
	\begin{equation}
		H(\hat{\vect k},x)=2v \left( \begin{array}{cccc}
			0 & \hat{\pi}^{3}_{+}& 0 & 0 \\
			\hat{\pi}^{3}_{-} & 0 & 0 & 0 \\
			0 & 0 & 0 & -\hat{\pi}^{3}_{+} \\
			0 & 0 & -\hat{\pi}^{3}_{-} & 0
		\end{array}\right).
	\end{equation}	
To solve the eigenvalue problem, we can separate variables and write the eigenspinors as plane waves in the $y$-direction. This is due to the fact that $[H(k), \hat{k}_{y}] = 0$ implies conservation of momentum along the $y$-direction $\Psi(x,y) = e^{ik_{y}y}\Psi(x)$ with
	$ \Psi(x)=\left[\Psi_{1}(x),\Psi_{2}(x),\Psi_{3}(x),\Psi_{4}(x)\right]^{T} $.
We may now define operators
	\begin{align}
	&	\hat{a}=\dfrac{1}{\sqrt{2 B}}\left[\hat{k}_{x}+i(k_{y}-B\hat x)\right],	
	\\
	&
		\hat{a}^{\dagger}=\dfrac{1}{\sqrt{2 B}}\left[\hat{k}_{x}-i(k_{y}-B\hat x)\right],	
  \label{eq:creation}
	\end{align}
that satisfy the commutation relation $[\hat{a},\hat{a}^{\dagger}]=\mathbb{I}$ and can therefore be interpreted as ladder operators. Using this notation, the Hamiltonian becomes
	\begin{equation}
			H=\omega_{c}\left( \begin{array}{cccc}
				0 & \hat{a}^{3}& 0 & 0 \\
				\hat{a}^{\dagger 3} & 0 & 0 & 0 \\
				0 & 0 & 0 & -\hat{a}^{3} \\
				0 & 0 & -\hat{a}^{\dagger 3} & 0
			\end{array}\right),
	\end{equation}
where $\omega_{c}= 2v\left(2e B\right)^{3/2} $ is called cyclotron frequency (recall $v$ is not a velocity). The eigenvalue problem  $H(k)\Psi(x,y)=E\Psi(x,y)$ for the second spinor $\Psi_{2}(x)$ component can be reduced to  
	\begin{equation}\label{h5}
		\omega^{2}_{c}\hat{a}^{\dagger3}\hat{a}^{3}\Psi_{2}(x)=E^{2}\Psi_{2}(x),
	\end{equation}
	which can be solved by harmonic oscillator states $\Psi_{2}(x)=\big|n\big>$. Associated  energies are given as	
	\begin{equation}\label{j1}
		E_{n,s}=s \omega_{c}\sqrt{n(n-1)(n-2)}, 
	\end{equation}
where  $ s=\pm 1$. Using as an ansatz $\Psi_{2}(x)=\big|n\big>$ in this equation, we can obtain full normalized spinors as
	\begin{align}
	&	\Psi_{n}^{(h)}(x,y) =\dfrac{1}{\sqrt{2}}\left( \begin{array}{c}
			s\big|n-3\big>  \\
		\big|n\big>\\
			0\\
			0
		\end{array}\right)e^{ik_{y}y},        \qquad     n\geq 3,
  \end{align}
  and using a similar relation for the fourth component, we find
	\begin{align}
	\Psi_{n}^{(-h)}(x,y) =\dfrac{1}{\sqrt{2}}\left( \begin{array}{c}
		0  \\
		0\\
		-s\big|n-3\big>\\
		\big|n\big>
	\end{array}\right)e^{ik_{y}y},        \qquad     n\geq 3,
\end{align}
Here, the label ${(h)}$ stands for spinors with contributions that come solely from the upper block and ${(-h)}$ for the lower block in the Hamiltonian.

A special case must be considered for describing the $n<3$ states since they require different normalization and do not come as pairs with opposing signs. For the case $m=0,1,2$, we find the spinors
	\begin{align}
		&\Psi_{l}^{(h)}(x,y) =\left( \begin{array}{c}
			0\\
			\big|l\big> \\
			0\\
			0
		\end{array}\right)e^{ik_{y}y},\quad l<3, \\
	&\Psi_{l}^{(-h)} (x,y) =\left( \begin{array}{c}
		0\\
		0 \\
		0\\
	\big|l\big>
	\end{array}\right)e^{ik_{y}y},\quad l<3
	\label{eq:zero_en_spinors},
\end{align}
where we find that the zero energy ground state is sixfold degenerate. This observation is interesting because it contrasts that of single-layer graphene, which has a zero energy state with no degeneracy. 
This ground-state degeneracy has interesting consequences and impacts the Hall conductivity. We begin our discussion of this with the typical formula for the Hall conductivity at finite temperature
\begin{equation}
    \sigma_{xy}=i\sum_{n, n'}\left(f\left(E^n\right)-f\left(E^{n'}\right)\right) \frac{\left\langle \psi_n\left|j_x\right| \psi_{n'}\right\rangle\left\langle \psi_{n'}\left|j_y\right| \psi_n\right\rangle}{\left(E^n-E^{n'}\right)^2},
\end{equation}
where $|\psi_n\rangle$ is any eigenstate of the Hamiltonian, $f(E)=(\exp(\beta(E-\mu))+1)^{-1}$ is the Fermi-Dirac distribution, $\mu$ the chemical potential, and $\beta$ the inverse temperture. Current operators are found in the usual way by taking the derivative of the Hamiltonian with respect to a vector potential. For our eigenstates and eigenvalues, we find a unitless reduced Hall conductivity $\sigma_{xy}^{(r)}$, which is given as
\begin{equation}
\begin{aligned}
    \sigma_{xy}^{(r)}&=\frac{h\sigma_{xy}}{4e^2}=\frac{9}{2}\omega_c^2\sum_{n,\sigma,\sigma^\prime}n(n-1)\frac{f({E_{n+1,\sigma}})-f({E_{n,\sigma^\prime}})}{(E_{n,\sigma^\prime}-E_{n+1,\sigma})^2}.
\end{aligned}
\end{equation}
Here, the sum is over $n\in \mathbb{N}$ and $\sigma,\sigma^\prime\in \{-1,1\}$.
The result of the computation can be seen in Fig. \ref{fig:Hall_conductivity_plot_kz0} below.
\begin{figure*}[htbp!]
    \centering
    \includegraphics[width=\linewidth]{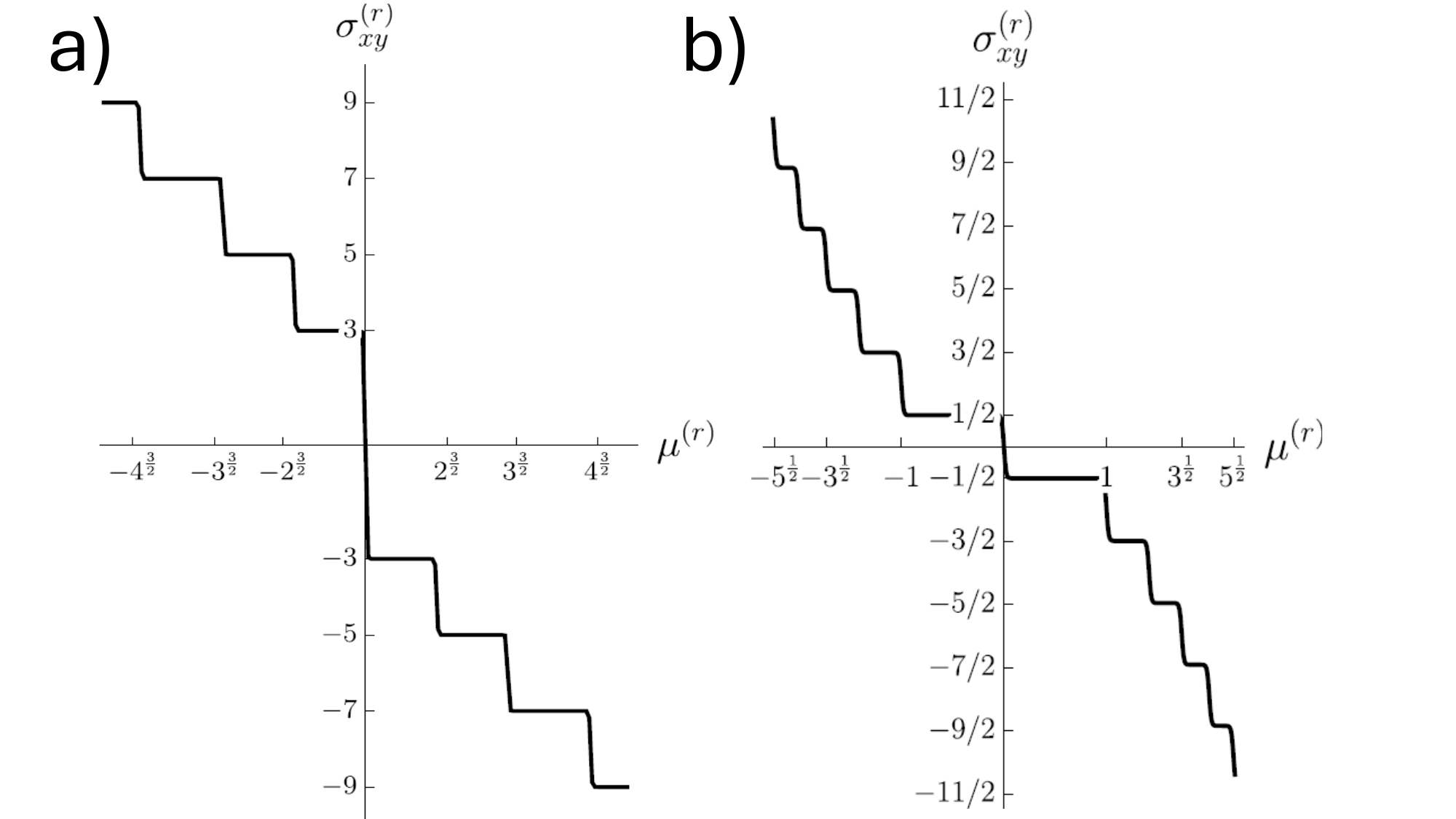} 
    \caption{Plot of the dimensionless Hall conductivity $\sigma_{xy}^{(r)}$ as function of dimensionless chemical potential $\mu^{(r)}=\mu/\omega_c$. Subfigure a) shows the Hall conductivity for our problem of a cubic Dirac semi-metal ($\beta=50/\omega_c$), and Subfigure b) the case of graphene ($\beta=100/\omega_c$). }
    \label{fig:Hall_conductivity_plot_kz0}
\end{figure*}
It shows that the step in Hall conductivity at zero chemical potential is $3/2$ times the step size of other steps in the quantized Hall conductivity. This result differs from the case of graphene, which we show for comparison where the first step is half the usual step size - for a typical Dirac semi-metal, the result has the same properties as graphene. The reason for this behavior can be found in the six-fold degenerate lowest Landau level of the cubic Dirac semi-metal, which has 3/2 the degeneracy of other states. This property contrasts graphene, where the lowest Landau level has half the degeneracy of other levels. This result also differs from a typical metal where all steps are equally spaced. The second observation one may make is that in graphene, steps occur at a dimensionless chemical potential $\mu^{(r)}=\sqrt{n}$. In contrast, in the cubic Dirac semi-metal case, they occur at values that asymptotically go to $\mu^{(r)}=n^{3/2}$. Both observations combined with the step height (it is twice that of a cubic Weyl semi-metal \cite{Barlas_2012}) - which is due to the level degeneracy of the cubic Dirac semi-metal's Landau levels - can be used to identify such materials via the quantized Hall conductivity experimentally.

\subsection{Quasi-2D isotropic case \texorpdfstring{$k_z\neq0$}{kz=0}}
\begin{figure}[H]
    \centering
    \includegraphics[width=1\linewidth]{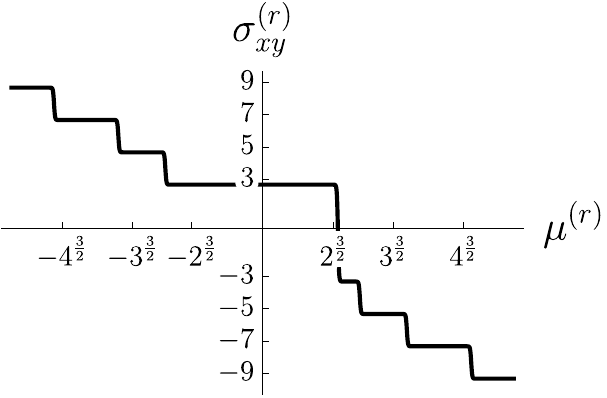}
    \caption{Plot of the dimensionless Hall conductivity $\sigma_{xy}^{(r)}$ as function of dimensionless chemical potential $\mu^{(r)}=\mu/\omega_c$ at inverse temperature $\beta=50/\omega_c$ and with "mass" term $m=-3$.}
    \label{fig:Hall_conductivity_plot_kz3}
\end{figure}

	After considering the case of semi-metal with $k_z=0$ (in addition to being isotropic in the $x$-$y$ plane), we now consider the case where the dimension of the material in $z$-direction is sufficiently large such that considering non-zero momentum $k_{z}\neq 0$ is not only relevant for high energies. As previously, we will restrict our discussion to the isotropic case. To keep our description consistent in units of the cyclotron frequency $\omega_c$, we define $m= \frac{v_{z} k_{z} } {2v\left(\sqrt{2B}\right)^{3}} $, which can be interpreted as a mass-like term in the original Hamiltonian model (it causes a gap in the excitation spectrum). The Hamiltonian then takes the form
	\begin{equation}
		H=\omega_{c}\left( \begin{array}{cccc}
			m & \hat{a}^{3}& 0 & 0 \\
			\hat{a}^{\dagger3} & -m & 0 & 0 \\
			0 & 0 & -m & -\hat{a}^{3} \\
			0 & 0 & -\hat{a}^{\dagger3} & m
		\end{array}\right).	
	\end{equation}
From here, finding the Landau levels is almost immediate and follows almost the same steps as in the previous section.
We find
	\begin{equation}
		E_{n,s}(m)=s \omega_{c}\sqrt{n(n-1)(n-2)+m^{2}}, \quad     n\geq 3,
		\label{eq:5-91}
	\end{equation}	
	with $s=\pm 1$, where
normalized eigenspinors are given as
	\begin{align}
	&	\Psi_{n,s}^{(h)}(x,y) =N\left( \begin{array}{c}
			sc_{n}	\big|n-3\big>\\
			\big|n\big>\\
			0\\
			0
		\end{array}\right)e^{ik_{y}y},\\
		&
	\Psi_{n}^{(-h)}(x,y) =N\left( \begin{array}{c}
		0\\
		0\\
		-sc_{n}\big|n-3\big>\\
		\big|n\big>
	\end{array}\right)e^{ik_{y}y},      
\end{align}
with the constants
\begin{align}
& c_n=\dfrac{E_{n,s}(0)}{E_{n,s}(m)-m}, 
\ \ N=\dfrac{1}{\sqrt{c_{n} ^{2} +1}}.
\end{align}
It is important to note that the case of $n<3$ again requires special care and has the same eigenspinors as given in Eq. \eqref{eq:zero_en_spinors} with corresponding eigenvalues
	\begin{equation}
	    E_n^{(\pm h)}=\mp \omega_cm,
	\end{equation}
	which are each thrice degenerate. It is important to stress that the "$-$" sign corresponds to the upper block of the Hamiltonian and the "$+$" sign to the lower block, which is unlike the $n>3$ result that has both signs in both cases.
	Much like the truly one-dimensional case, we can plot a Hall conductivity and find the result in Fig. \ref{fig:Hall_conductivity_plot_kz3} below
We observe that the effect of the mass term $m$ is that some plateaus have been broadened in width while others have been shrunk. This behavior would allow experimentalists to recognize if electrons carry momentum in the z-direction.

	\section{Perturbative energy levels for the case \texorpdfstring{$v_{y}\neq v_{x}$}{vy=v}}\label{pp3}
	While it was easy to find exact energy levels in the isotropic case of $v_{y}=v_{x}$, this task is not simple using analytical means for the anisotropic case of $v_{y}\neq\ v_{x}$. Introducing an anisotropy parameter $\lambda=\frac{1-v_y/v_x}{2}$ is useful. This approach allows us to write the Hamiltonian in a convenient form
\begin{equation}
	H=\omega_{c}( h_{0}+\lambda h_{1})\otimes \sigma_z,
\end{equation}	
with $\lambda$ as perturbative parameter and definitions
\begin{align}
	h_{0}=\left( \begin{array}{cc}
		0& \hat{a}^{3} \\
		\hat{a}^{\dagger 3}& 0
	\end{array}\right),\ h_{1}= \left( \begin{array}{cc}
	0&  \hat{a}^{\dagger 3}-\hat{a}^{3}\\
	\hat{a}^{3}-\hat{a}^{\dagger 3}& 0
\end{array}\right).
\end{align}
With this, $h_{0}$ corresponds to an exactly solvable isotropic Hamiltonian, and $ \lambda h_{1}$ is considered a perturbation. 
  Eigenvalues $E^{0}_{n}$ are non-degenerate in the case $n\geq 3$ and corrections to the energy to first order can be computed straightforwardly by the standard expressions for Reighleigh-Schr\"odinger perturbation theory \cite{https://doi.org/10.1002/andp.19263851302} as 
\begin{align}
	E_{n} 
	=E_{n}^{0}\left( 1-\lambda\right) 
	\label{eq:en_lowest_correct}.
\end{align}	
Like in previous sections, the degenerate zero eigenvalues case requires special care and must be treated using degenerate perturbation theory. Interestingly, first-order degenerate perturbation theory yields no correction such that the expression \eqref{eq:en_lowest_correct} that includes the first-order corrections to energies above remains valid in all cases. Therefore, it is clear that a small anisotropy is not expected to have much of an impact on the qualitative picture of the Hall conductivity (to leading order $\omega\to (1-\lambda)\omega_c$ is the only modification).

	\section{Semi-classical treatment}\label{pp4}
	Since a completely quantum mechanical treatment for the anisotropic case is challenging, we will resort to a semi-classical treatment in this section. This approach allows us to gain further insights, such as the shapes of particle trajectories under the influence of a magnetic field, and further insights into the anisotropic regime, which we stress is difficult to access in a complete quantum treatment.
	\subsection{Review of Bohr-Sommerfeld-type semiclassics for matrix Hamiltonians}
	A general and powerful method to determine energy levels by studying classical trajectories is the so-called Gutzwiller approach to the semi-classical density of states \cite{10.1063/1.1665596}. This approach has been generalized to arbitrary matrix-valued Hamiltonians in \cite{vogl2017semiclassics}. While the method developed in \cite{vogl2017semiclassics} is quite general and allows for treatments of various special cases, our current problem of determining Landau levels for a cubic Dirac semi-metal is much simpler. Notably, we have three significant simplifications:
	\begin{enumerate}
	    \item It is enough to restrict ourselves to one-dimensional Hamiltonians. We may do so because, in our case of Landau levels, $k_y$ and $k_z$ can be treated as parameters entering a 1D Hamiltonian - that is, as complex numbers rather than bonafide operators.
	    \item The eigenvalues of the classical limit Hamiltonian $H(p,x)$, where the momentum operator $\hat p$ was mapped to a complex number $p$, are non-degenerate. This simplification is possible because the $h(k)$ blocks in equation \eqref{eq:hfull} are not coupled, and it is, therefore, enough to consider them separately, and their eigenvalues can be found to be non-degenerate.
	    \item We may ignore degeneracy factors that arise from different Landau orbits over the plane. This simplification is possible because they only enter the density of states and not in expressions for the energy levels that interest us.
	\end{enumerate}
These three simplifications mean that in what follows, we may directly employ the modified Bohr-Sommerfeld quantization condition that was derived in \cite{vogl2017semiclassics} and is given below
	\begin{equation}
	2S_{\alpha}-\hbar\left[\pi(\nu_{\alpha}+4n)-2\int_0^{T_{\alpha}}dt M_\alpha\right]=0.
	\label{eq:bohr-sommerfeld}
	\end{equation}
	In what follows, we will choose units with $\hbar=1$.
We now move on to explain the different terms that enter this expression.
	Hereby, $S_\alpha$ is the classical action for a primitive periodic orbit - an orbit in phase space that has only been traversed once. The index $\alpha$ accounts for different periodic orbits being possible at a fixed energy. There are two ways this can happen.
	\begin{enumerate}
	    \item Potentials can allow for different periodic orbits at a given energy $E$ like in Fig. \ref{fig:two-periodic_orbits} below.
	\begin{figure}[H]
	\centering
	\includegraphics[width=0.7\linewidth]{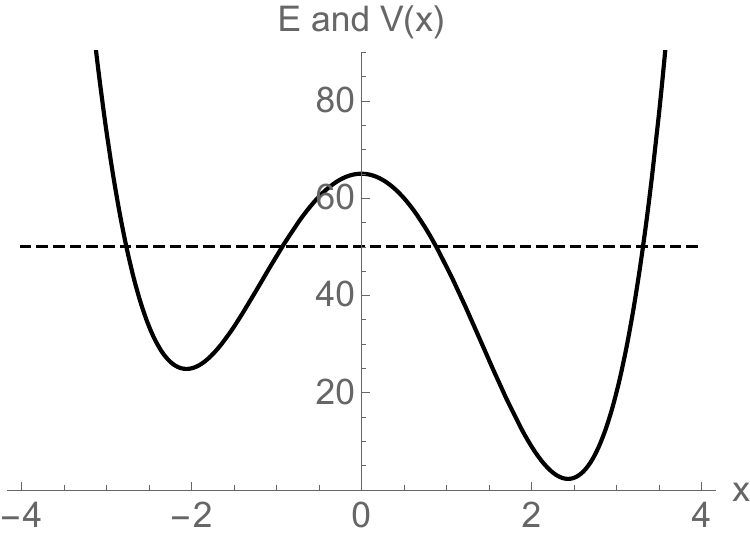}
	\caption{Plot of a potential landscape $V(x)$ (thick black line) and energy as a dashed line. Two distinct potential pots are visible. Each will have its own associated classical periodic orbits with an action $S_\alpha$.}
	\label{fig:two-periodic_orbits}
	\end{figure}
	\item We deal with matrix Hamiltonians. That is, to the lowest order, the Schr\"odinger equation is replaced by a simple matrix equation
\begin{equation}
    H(x,p_x=\partial_x S)V=EV.
\end{equation}
For an $n\times n$ Hamiltonian this can be diagonalized to determine $n$ eigenvectors $V_i$ and $n$ eigenvalue equations that determine actions $S_\alpha$
\begin{equation}
    H_i(x,p_x=\partial_x S_\alpha)=E,
\end{equation}
	which are ordinary Hamilton-Jacobi equations and $H_i$ are eigenvalues of $H(x,\partial_x S)$. These Hamilton-Jacobi equations can have one or more periodic orbits associated with a given energy $E$. 
	\end{enumerate}
	Next, in our expression, Eq. \ref{eq:bohr-sommerfeld} enters $\nu_\alpha$, which is the so-called Maslov index which for 1D motion counts the number of reflections by a potential barrier (or more generally directional changes) that a particle experiences while it traverses a periodic orbit. The last term that enters is the so-called semi-classical phase factor
	\begin{equation}
	    M_\alpha=\mathrm{Im}(V_\alpha^\dag [\partial_{p_x} H(x,p_x)]\partial_x V_\alpha),
	    \label{eq:semiclass_berry}
	\end{equation}
	which bears some resemblance to a Berry phase as discussed in \cite{PhysRevB.77.245413,KEPPELER200340,BOLTE1999125,PhysRevLett.81.1987,vogl2017semiclassics}. Hereby, $V_\alpha$ is the eigenvector of the classical Hamiltonian matrix $H(x,\partial_x S_\alpha)$ corresponding to the orbit $\alpha$.
	
	\subsection{The Hamiltonian system}\label{pp5}
	
For our computations of semi-classical energies, we will consider the upper block of \eqref{eq:hfull} only, that is, Eq. \eqref{eq:hupperblock} with the magnetic field introduced as before via minimal substitution. This simplification is possible because the second block in equation \eqref{eq:hfull} will yield (up to a sign) the same energies as the other block. The semi-classical limit of the Hamiltonian \eqref{eq:hupperblock} is then obtained by replacing momentum operators with canonical momenta $\hat p_i\rightarrow \partial_{x_{i}} S=p_i$. We obtain the classical expression
\begin{equation}
	h(\pi_i)=2v_x\begin{pmatrix}
	    0&\pi_+^3+\lambda(\pi_-^3-\pi_+^3)\\
	    \pi_-^3+\lambda(\pi_+^3-\pi_-^3)&0
	\end{pmatrix},
\end{equation}
and we introduced the shorthand notations
\begin{equation}
\begin{aligned}
	& \pi_{\pm}=p_{x}\pm i(p_{y}+eBx),
	\end{aligned}
	\label{1230}
\end{equation} 
where  $p_{i}$ are canonical momenta and $\pi_i$ are kinetic momenta. The parameter $\lambda=\frac{1-v_y/v_x}{2}$ measures the degree of anisotropy as discussed previously.

\subsection{Isotropic case \texorpdfstring{$k_{z} = 0 $}{kz=0}}	\label{pp6}
In the isotropic case, we may set $\lambda=0$ and $v_x=v_y=v$ and for a thin sheet, low energies correspond to $k_z=0$ to find
\begin{equation}
h(\pi)=2 v \left( \begin{array}{cc}
	0 & \pi^{3}_{+} \\
	\pi^{3}_{-} & 0 \\
\end{array}\right).
\end{equation}
The eigenvalues
 are obtained from	the determinant as
		$ 	\det [h(\pi_i)-E]=0 $
	\begin{equation}\label{aa11}
		\begin{split}
			E_s=&2sv\sqrt{(\pi_{+}\pi_{-})^{3}}
		\end{split}, 	
	\end{equation}
 where $s=\pm 1$.
 The corresponding normalized eigenvectors are given by
 \begin{equation}
 V=\dfrac{1}{\sqrt{2}}
  \begin{pmatrix}
 	 \frac{2v}{E_s}\pi_+^{3}  \\
 	1  
 \end{pmatrix}.
\end{equation}
 At this level, we see that there is two classical Hamiltonians of the form
 \begin{equation}
H(p_i,x) =2sv\left( p_x ^{2}+\left(  p_y +eBx\right)^{2} \right)^{3/2}.
\label{eq:ham_class_isotrop}
\end{equation}	
\subsubsection{Time for an orbit}
%
%
Another ingredient that enters our expression for the generalized Bohr-Sommerfeld quantization condition is the time for an orbit. Here, we start by investigating the equations of motion, which are given by Hamilton's equations
	\begin{equation}\label{hameq}
		\dot{q}=\dfrac{\partial H}{\partial p}, \quad 
		\dot{p}=-\dfrac{\partial H}{\partial q},
	\end{equation}
with $q$  is the position and $p$ is  momentum coordinates. 
In our case, we find
\begin{equation}
	\begin{aligned}
&			\dot{x}= 3s\left(4|E|v^2\right)^{\frac{1}{3}}  p_{x},\\
&
			\dot{y}=3s\left(4|E|v^2\right)^{\frac{1}{3}}  (p_y + eB x),\\ 
			&
			\dot{p}_{x}=3seB\left(4|E|v^2\right)^{\frac{1}{3}}  (p_y + eB x),
	\\
		&\dot{p}_{y}=0,
		\end{aligned}
		\label{a6}
		\end{equation}

From \eqref{a6} one finds (taking a time derivative of $\dot x$ and using the equation for $\dot p_x$) that the equation of motion for the $x$-coordinate is given as
\begin{equation}
	\ddot x= -\omega^2 (x+x_s)
	\label{eq:HO},
\end{equation}
which constitutes a shifted harmonic oscillator with position shift $x_s=p_y/(eB)$ (note that $p_y$ is constant as shown above). It is important to note that this equation is valid for both classical Hamiltonians in Eq. \eqref{eq:ham_class_isotrop}. It also means that it is enough to consider only one of the eigenvalues to find semi-classical energy levels. Indeed, in the remainder of this section, we will focus on only one of the Hamiltonians and, therefore, drop all indices related to orbits.
The oscillator frequency is found to be
\begin{equation}
    \omega=3(4|E|v^2)^{\frac{1}{3}}eB.
    \label{eq:osc_frequ_isotropkz0}
\end{equation}
The time for an orbit as per definition $\omega =2\pi/T$ then is then found to be
	\begin{equation}
		T=\dfrac{2\pi }{3(4|E|v^2)^{\frac{1}{3}}eB}.
	\end{equation} 
\subsubsection{Classical orbit, action, and Maslov index}
Now, from Eq. \eqref{eq:HO}, we find the solutions 
\begin{align}
& x=x_s+x_0\sin(\omega t),
&	p_{x}=seBx_0\cos\left(\omega t\right),
\label{solixma}
\end{align}
with amplitude 
\begin{equation}
    x_0=\frac{1}{eB}\left(\frac{|E|}{2v}\right)^{\frac{1}{3}},
\end{equation}
 which can be found for a given energy by placing the solutions \eqref{solixma} in Eq. \eqref{aa11}.
  Plotted, we get the typical phase space curve of a harmonic oscillator that can be seen below
 
 \begin{figure}[H]
	\centering
	\includegraphics[width=0.7\linewidth]{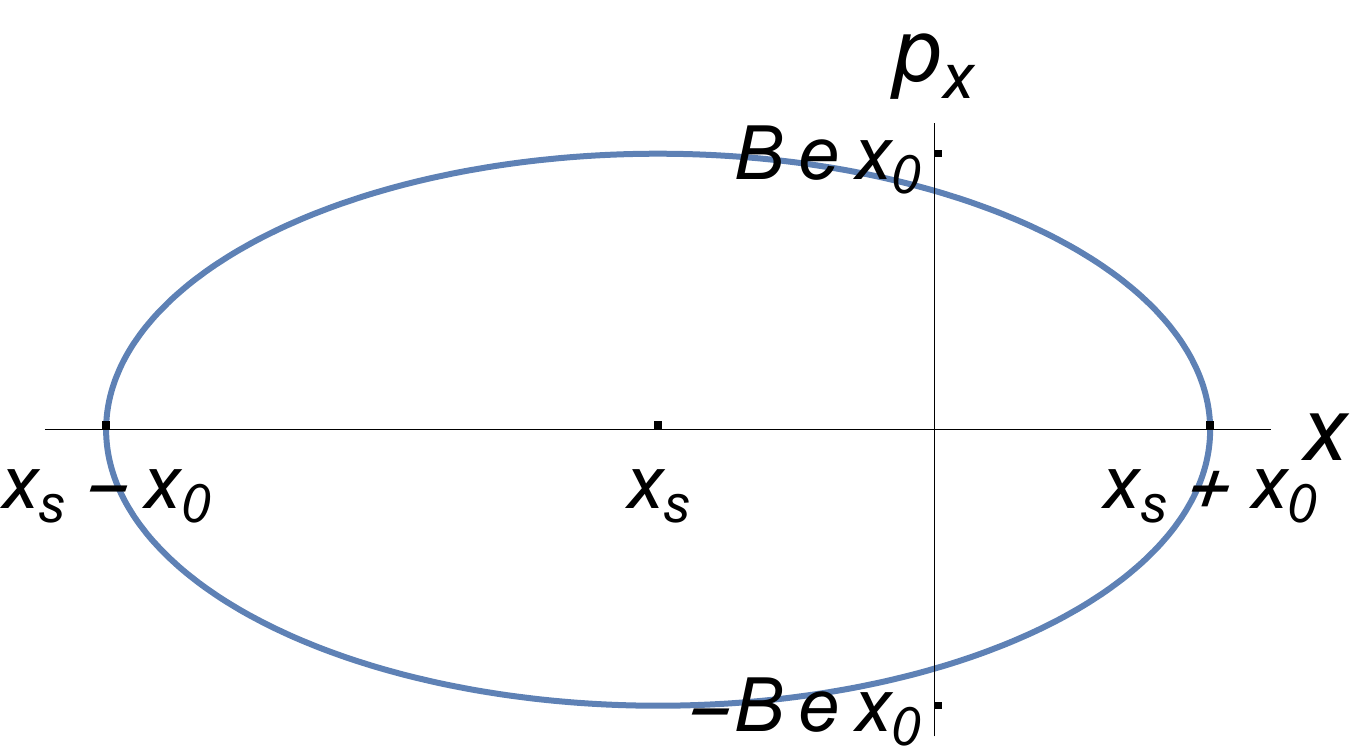}
	\caption{Phase space curve of an electron in cubic semi-metal subjected to a constant magnetic
		field.}
	\label{fig:px-en-x}
\end{figure}
\noindent which is just an elliptic orbit. This curve directly tells us that the particle along each orbit changes direction twice, and therefore, the Maslov index is 
\begin{equation}
    \nu_\alpha=2
    \label{eq:isotrop_case_maslov}.
\end{equation}
We may next use  $p_x=\partial_x S$ in Eq. \eqref{aa11} to find the action for an orbit as
\begin{equation}\label{m0}
	\begin{split}
		S&=\int \sqrt{\left(\frac{|E|}{2v}\right)^{\frac{2}{3}}-(p_y+eBx)^2}dx\\
	\end{split}.
\end{equation}
This integral for a single period of a primitive orbit gives
\begin{equation}
    S=\frac{\pi}{e B} \left(\frac{|E|}{2v}\right)^{\frac{2}{3}}
    \label{eq:class_action_isotropic}.
\end{equation}

\subsubsection{Semi-classical Berry-like phase and energy levels}	
For the semi-classical phase term using \eqref{eq:semiclass_berry} we find
	\begin{align}
	 M_{\pm}=9eB\left(\frac{|E|v^2}{2}\right)^{\frac{1}{3}},	
	 \label{eq:berry_like_isotropic}
\end{align}		
which is a constant, and therefore the integral $\int_0^Tdt M=MT=3\pi$ is trivial.

We may use our results \eqref{eq:class_action_isotropic}, \eqref{eq:isotrop_case_maslov}, \eqref{eq:berry_like_isotropic} and put them into \eqref{eq:bohr-sommerfeld} to find that our semiclassical energies are then given by
\begin{equation}
    E=\pm \omega_c(n-1)^{3/2},
\end{equation}
with $\omega_{c}= 2v\left(2e B\right)^{3/2} $.

	\subsubsection{Performance of semi-classical results}
Understanding under what conditions the semi-classical results perform well compared to the exact results obtained earlier is helpful. Of course, we see directly that the results are different. This observation poses the question under which conditions the results agree.

Typically, semi-classical results are reliable for large quantum numbers. Therefore, we do an expansion for large $n$ of both the exact $E_\mathrm{exact}$ and semi-classical $E_\mathrm{semcl}$ results to find
\begin{equation}
    \hspace{-0.3cm}\begin{aligned}
    &E_\mathrm{exact}=\pm\omega_c\left(n^{3/2}-\frac{3 }{2}n^{\frac{1}{2}}+\frac{3 }{8}n^{-\frac{1}{2}}\right)+\mathcal{O}\left(n^{-\frac{3}{3}}\right),\\
    &E_\mathrm{semcl}=\pm\omega_c\left(n^{3/2}-\frac{3 }{2}n^{\frac{1}{2}}-\frac{1 }{8}n^{-\frac{1}{2}}\right)+\mathcal{O}\left(n^{-\frac{3}{3}}\right),
    \end{aligned}
\end{equation}
 where we can see that results agree well in the limit of large quantum numbers - they only disagree at order $\mathcal{O}(n^{-1/2})$ as it is typical in the semi-classical approach - because typically larger quantum numbers correspond to a larger classical action.
To better understand how accurate the approximation is, we plot the density of states below and energy levels as a function of $n$. The approximation's effectiveness is also visualized in the figure below.
\begin{figure}[H]
    \includegraphics[width=0.5\linewidth]{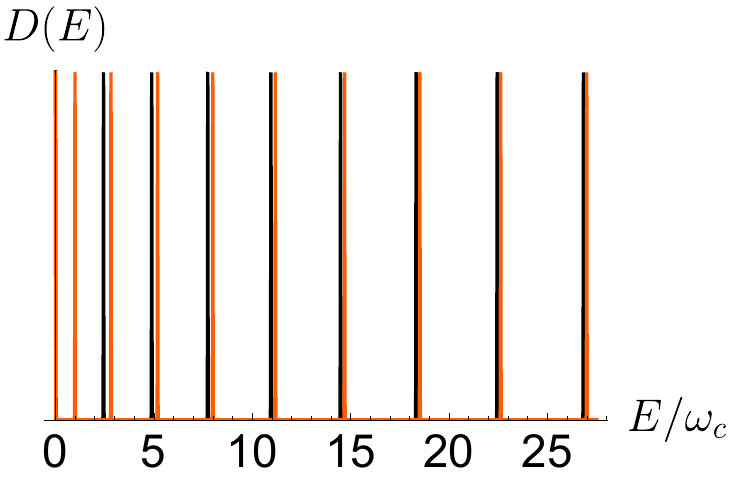}\includegraphics[width=0.5\linewidth]{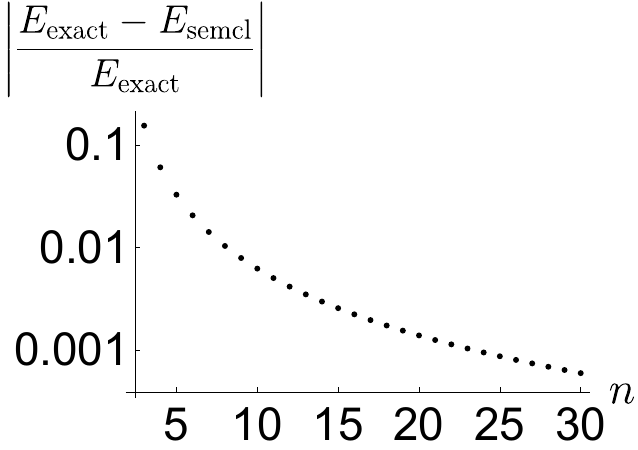}
    \caption{(color online) The left column of the figure shows a plot of the density of states, and the right column shows the relative error between exact and approximate energies.}
    \label{fig:my_label}
\end{figure}
The semi-classical energy levels agree very well with exact results except for the lowest energy. Furthermore, as expected, the approximation error decreases rapidly as a function of the quantum number $n$.

\subsection{Quasi-2D isotropic limit }\label{pp7}
Next, we consider the case $p_z\neq 0$, which can describe a sheet of finite thickness. To obtain a simpler notation that we can compare to the exact quantum case, we set $p_z=m\omega_c/(2v_z)$. The notation is suggestive and clarifies that $p_z$ takes the role of a mass term $m$. The Hamiltonian then takes the form
\begin{equation}
h(\pi)=2 v \left( \begin{array}{cc}
	\frac{m\omega_c}{2v} & \pi^{3}_{+} \\
	\pi^{3}_{-} & -\frac{m\omega_c}{2v}  \\
\end{array}\right).
\end{equation}
The eigenvalues are readily found as
\begin{equation}\label{j1j}
E=2vs\sqrt{(\pi_{+}\pi_{-})^{3}+\left(\frac{m\omega_c}{2v}\right)^2},
\end{equation}
with $s$ the sign of the energy. The corresponding eigenvectors can be written as
  \begin{align}
  	V=\sqrt{2}\sqrt{1-\frac{m\omega_c}{E}}\begin{pmatrix}
  		\frac{ \pi_+^3 v}{E-2 m\omega_c}\\1
  	\end{pmatrix}.
\end{align} 
The classical system will reduce to a harmonic oscillator after using the same methods as in the previous section. However, this time with a slightly modified driving frequency
\begin{equation}
    \omega =3\left(\frac{2v\left(\left|E|^2-m^2\omega_c^2\right.\right)}{|E|^{3/2}}\right)^{\frac{2}{3}}eB.
\end{equation}
Considering the limit of small mass and the relation to the frequency in the massless case, Eq. \eqref{eq:osc_frequ_isotropkz0} is interesting. We will use the term $\omega_0$ to refer to the frequency of the massless case below. In the limit of small masses, the oscillation frequency becomes
\begin{equation}
    \omega\approx\omega_0-\frac{2}{3}\frac{m^2\omega_c^2}{|E|^2}\omega_0,
\end{equation}
which tells us that a mass term reduces the oscillation frequency just as intuition would dictate. Because we have harmonic motion, we find a Maslov index 
\begin{equation}
    \nu=2.
\end{equation}
The action of a primitive can be computed in analogy to the previous case and is given as
\begin{equation}
    S=\frac{\pi  (|E|^2- m^2\omega_c^2)^{\frac{1}{3}}}{ 2^{2/3}eB v^{2/3}}.
\end{equation}
Lastly, we find that the semi-classical phase factor has the form
\begin{equation}
    M=\frac{9 B e \left(v \left(|E|^2-4 m^2\omega^2/v^2\right)\right)^{2/3}}{2^{\frac{1}{3}} E},
\end{equation}
 which is a constant. Therefore, the integral $\int_0^Tdt M=2\pi M/\omega=3\pi$ is trivial.

Applying the Bohr-Sommerfeld quantization condition Eq. \eqref{eq:bohr-sommerfeld} as before, we find that energy levels are given as
\begin{equation}
    E_\text{semcl}= \omega_c\sqrt{m^2+ (n-1)^3}.
\end{equation}
We may now compare the semi-classical result $E_\text{semcl}$ to the exact result Eq. \eqref{eq:5-91}, which we title $E_\text{exact}$ to find
\begin{equation}
    E_\text{semcl}-E_\text{exact}=(n-1)\frac{\omega_c}{2m}+\mathcal{O}(m^{-3}),
\end{equation}
which tells us that, similar to expectations, a mass term turns the energy levels more classical (the approximation error shrinks for large mass).

\subsection{Maximally anisotropic case $v_x=0$}
\label{sec:max_anisotrop}
\begin{figure}[H]
     \centering
     \includegraphics[width=1\linewidth]{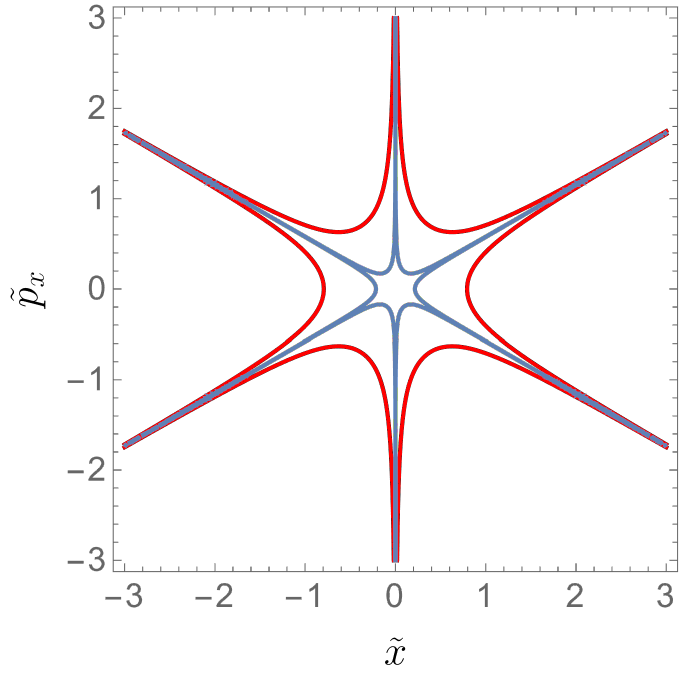}
     \caption{Plot of phase space trajectories in terms of unitless momentum $\tilde p_i=p_i(eB)^{-1/2}$, unitless energy $\epsilon=Ev^{-1}(eB)^{-3/2}$ and position $\tilde x=x(eB)^{1/2}$ (recall $v$ is not a velocity and we set $\hbar=1$). In both cases, we set $p_y=0$ because this term only leads to a shift of the trajectory center along the $x$-axis. The red curve is for a value $\epsilon=1$ and the blue one for a value $\epsilon=0.02$}
     \label{fig:star_orbit}
 \end{figure}

In the last part of the section, we consider a case that is difficult to solve by analytical means in an exact quantum treatment - the case of a maximally anisotropic cubic Dirac semi-metal. Maximum anisotropy is achieved for $v_x=0$ and $v_y=v$, where the classical Hamiltonian takes the form
\begin{equation}
    H=v\begin{pmatrix}
        0&\pi_-^3-\pi_+^3\\
        \pi_+^3-\pi_-^3&0
    \end{pmatrix}.
    \label{eq:aniso_trop_ham}
\end{equation}
 For this case, it is advantageous to combine Hamilton-Jacobi equations for different eigenvalues into a single Hamilton-Jacobi equation $\det(E-H)=0$, and one finds
 \begin{equation}
     E^2+v^2 \left(\pi_-^3-\pi_+^3\right)^2=0.
     \label{eq:tar_orbit-comb1}
 \end{equation}
 Eigenvectors that correspond to different solutions for $E$ are given as
 \begin{equation}
     V_\pm=\frac{1}{\sqrt{2}}\begin{pmatrix}
         \pm1\\1
     \end{pmatrix},
 \end{equation}
 and are independent of momenta and positions, which directly lets one realize that the semi-classical phase $M$ will have no contributions.
 It is, therefore, important to recognize that the main contribution to orbits will be from the classical phase space orbit. The combined Hamilton-Jacobi equation \eqref{eq:tar_orbit-comb1} can be interpreted as an implicit relation for the orbit in phase space
 \begin{equation}
    E^2-4 v^2 (B e x+p_y)^2 \left((B e x+p_y)^2-3 p_x^2\right)^2=0,
 \end{equation}
 which leads to an interesting phase space orbit shown in Fig. \ref{fig:star_orbit}.
We observe that semi-classical particles in this problem traverse star-shaped orbits in phase space. These orbits do not close except at infinity (for a non-zero value of $v_x$, the closure happens at a finite value, though). It is clear from the figure that despite orbits that only close at infinity, the phase space volume enclosed by the trajectory is finite. This observation is interesting and peculiar because the finite phase space volume implies discretized energy levels corresponding to localized states. At the same time, the classical particles are not localized - they move to infinity along the jags of the star. This observation demonstrates how a particle can be localized for a quantum mechanical problem. The classical dynamics do not reflect this in the real space trajectory but only in phase space through a finite volume. The Maslov index can also easily be found from the figure if we note that barrier reflections happen when the momentum derivative becomes infinite - this happens at four spots along the orbit (the two spikes at $\tilde x=0$ and the indents at $\tilde p_x=0$). Therefore, the Maslov index in equation \eqref{eq:bohr-sommerfeld} is $\nu=4$.

The last ingredient in equation \eqref{eq:bohr-sommerfeld} to find semi-classical energies is the action, which is the volume enclosed by the orbit. The simplest way to do this is to realize that it is enough to compute the area of one of the four star sectors, which is marked in blue Fig. \ref{fig:star_sector} below.
\begin{figure}[H]
    \centering
    \includegraphics[width=1\linewidth]{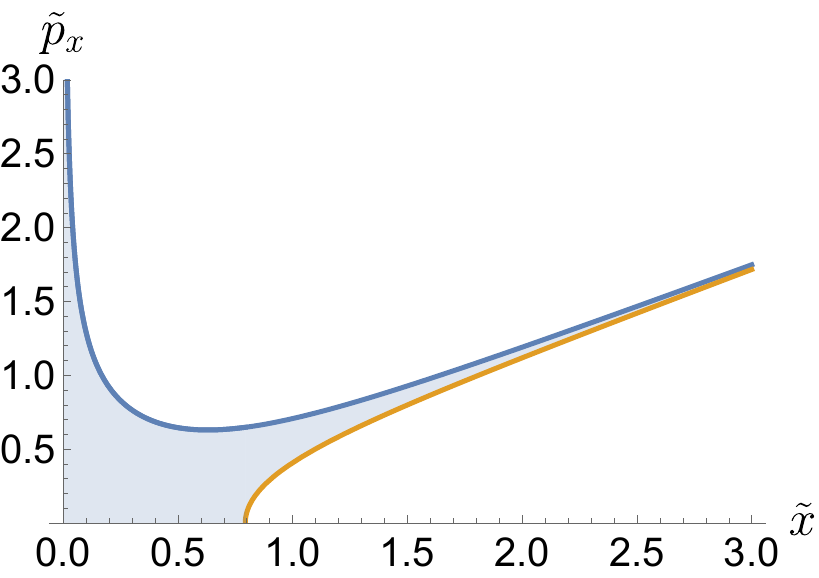}
    \caption{Plot of one sector of the star orbit. Marked in blue is the area one needs to compute. The blue curve is given by $\tilde p_x=\sqrt{x^2/3+\epsilon/(6 x)}$ and the orange curve by $\tilde p_x=\sqrt{x^2/3-\epsilon/(6 x)}$}
    \label{fig:star_sector}
\end{figure}
We find that the area can be separated into different pieces, giving a unitless action $s$ with 
\begin{equation}
    s/4=\int_0^{\infty} dx \sqrt{\frac{2 x^3+\epsilon}{6x}}-\int_{\epsilon^{1/3}2^{-1/3}}^\infty dx \sqrt{\frac{2 x^3-\epsilon}{6x}}.
\end{equation}
The integral simplifies considerably to give
\begin{equation}
    s=\frac{2^{4/3} \sqrt{\pi } \epsilon^{2/3} \Gamma \left(\frac{7}{6}\right)}{\Gamma \left(\frac{5}{3}\right)},
\end{equation}
where $\Gamma$ is the Gamma function.
After units are reintroduced, the result is inserted into the Bohr-Sommerfeld quantization condition, and the resulting relation solved for the energy. We find the semi-classical energy levels to be given by
\begin{equation}
    E_n=\hbar\omega_c\frac{\pi ^{3/4} \Gamma \left(\frac{5}{3}\right)^{3/2}}{8 \Gamma \left(\frac{7}{6}\right)^{3/2}}n^{3/2},
\end{equation}
which much like the isotropic case behave as $E_n\propto n^{3/2}$ just with a different proportionality constant. Of course, it is interesting that we obtained relatively simple closed-form results semiclassically, where a full quantum treatment is difficult. Even a proper numerical treatment is numerically expensive because particles are only weakly localized (for details, see appendix \ref{app:app1}). We compare our closed-form semi-classical approximation to the exact numeric results below to find out if our approximation is reliable.
\begin{figure}[H]
    \includegraphics[width=0.5\linewidth]{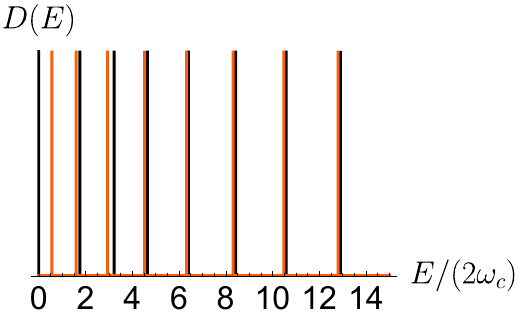}\includegraphics[width=0.5\linewidth]{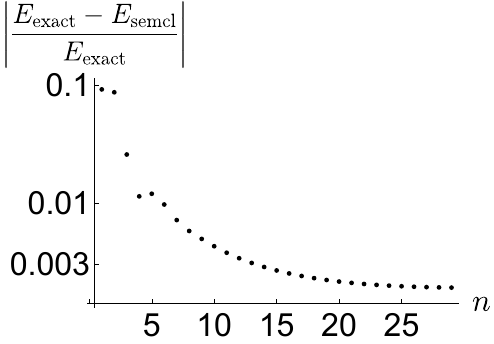}
    \caption{(color online) The left column of the figure shows a plot of the density of states, and the right column shows the relative error between exact and approximate energies.}
    \label{fig:my_label}
\end{figure}
The results are in excellent agreement with a complete numerical treatment. Therefore, we can see that the classically non-localized particles are weakly quantum localized. Moreover, the functional form of the energies tells us that even in an anisotropic case, one can expect similar behavior for the Hall conductivity as we have predicted for the isotropic case.

\section{Conclusion}
\label{sec:conclusion}
In summary, we have shown that the Landau levels of a cubic Dirac semi-metal exhibit
intriguing features both quantum mechanically and from a classical perspective. We have shown in a fully quantum mechanical treatment that one would expect experimentally detectable signatures of a cubic Dirac semi-metal in the Hall conductivity, and we have found explicit analytical expressions. While the anisotropic case could not be understood by employing a full quantum treatment, we made considerably much more progress on the semiclassical side.

Interestingly, we found that in the anisotropic case, the electrons do not localize classically. Rather they move infinitely far from the origin. However, they were found to be quantum mechanically weakly localized, giving rise to discrete energy levels. Our work also provides an exciting example of quantum corrections playing a significant role in localization, enriching the phenomenological literature in this respect. Nevertheless, weak localization from the classical side has left its imprint in that the numerics for finding exact quantum energy levels are relatively expensive when using a local set of basis functions.

\section{Acknowledgements}
The authors gratefully acknowledge the support provided by the Deanship of Research Oversight and Coordination (DROC) at King Fahd University of Petroleum \& Minerals (KFUPM) for funding this work through exploratory research grant No. ER221002.

\bibliography{./bibliography}

\appendix
\section{Numerical solution of the maximally anisotropic case}
\label{app:app1}
For a numerical treatment, we have expressed the Hamiltonian \eqref{eq:aniso_trop_ham} in terms of the usual Landau basis as
\begin{equation}
    h=\frac{\omega_c}{2}\begin{pmatrix}
        0&{a^{\dag}}^3-a^3\\
        {a}^3-{a^{\dag}}^3&0
    \end{pmatrix},
\end{equation}
and we used that we may write
\begin{equation}
    (a^3)_{ij}=\sqrt{2i+3i^2+i^3}\delta_{i+3,j}.
\end{equation}
A naive expansion letting $i$ and $j$ run from zero to a cut-off does not yield valid results. Indeed, as one can already check in the isotropic case, such an approach leads to issues with unphysical states at zero energy. One finds a sixfold fold degeneracy instead of the actual physical threefold degeneracy. Luckily, there is a simple way to fix this. If one replaces
\begin{equation}
    a^3\to (a^3)_{\text{chop,x,y}},
\end{equation}
where $(a^3)_{\text{c,x}}$ has the last three columns chopped from the matrix and $(a^3)_{\text{c,y}}$ has the last three rows chopped. In the isotropic case, this replacement leads to the correct number of zero modes and the correct eigenvectors for the zero modes, which justifies this replacement.

For the anisotropic case, we may then write the numerical Hamiltonian as
\begin{equation}
    h_\text{chop}=\frac{\omega_c}{2}\begin{pmatrix}
        0&({a^{\dag}}^3)_{\text{c,x}}-(a^3)_{\text{c,x}}\\
        ({a}^3)_{\text{c,x}}-({a^{\dag}}^3)_{\text{c,y}}&0
    \end{pmatrix},
\end{equation}
which leads to the results employed in the main text. We should also note that this approach is slowly converging, so a basis of $\sim 2^{22}$ states was needed (close to state-of-the-art system sizes in exact diagonalization) to generate results from the main text. The resulting Python code employing sparse matrix diagonalization is available from the authors upon reasonable request.
\end{document}